\documentclass[3p,times,procedia]{elsarticle}
\usepackage{nupha_ecrc}

\usepackage{graphicx}
\usepackage{wrapfig}
\usepackage[rightcaption]{sidecap}

\volume{00}

\firstpage{1}

\journalname{Nuclear Physics A}

\runauth{Alice Ohlson for the ALICE Collaboration}

\jid{nupha}

\jnltitlelogo{Nuclear Physics A}

\usepackage{amsmath}
\usepackage{amssymb}

\usepackage{lineno}

\usepackage[figuresright]{rotating}

\newcommand{\slfrac}[2]{\left.#1\right/#2}
\newcommand{\pta}{p_{\rm{T}}}
\newcommand{\pts}[1]{p_{\rm{T},#1}}

\newcommand{\anti}[1]{\overline{#1}}
\newcommand{\GeVc}{\text{ GeV/}c}

\begin{document}

\begin{frontmatter}



\dochead{XXVIIth International Conference on Ultrarelativistic Nucleus-Nucleus Collisions\\ (Quark Matter 2018)}

\title{Investigating correlated fluctuations of conserved charges with net-$\Lambda$ fluctuations in Pb--Pb collisions at ALICE}


\author{Alice Ohlson for the ALICE Collaboration}

\address{Physikalisches Institut, Ruprecht-Karls-Universit\"{a}t Heidelberg, Germany}

\begin{abstract}
Event-by-event fluctuations of conserved charges -- such as electric charge, strangeness, and baryon number -- in ultrarelativistic heavy-ion collisions provide insight into the properties of the quark-gluon plasma and the QCD phase diagram.  They can be related to the higher moments of the multiplicity distributions of identified particles, such as the $\Lambda$ baryon which carries both strangeness and baryon number and is thus of particular interest.  We present the first measurement of net-$\Lambda$ fluctuations in Pb--Pb collisions at $\sqrt{s_{\mathrm{NN}}} = 5.02$~TeV as a function of centrality and the pseudorapidity acceptance of the measurement.  The results are compared to expectations of the effects of global baryon number conservation as well as to predictions from the HIJING Monte Carlo event generator.  In this analysis the Identity Method is applied in a novel way to account for the combinatoric background in the invariant mass distribution.  
\end{abstract}

\begin{keyword}
heavy-ion collisions \sep quark-gluon plasma \sep fluctuations

\end{keyword}

\end{frontmatter}


\section{Fluctuations of conserved charges in heavy-ion collisions}

Within the Grand Canonical Ensemble framework, the event-by-event fluctuations of conserved quantities are related to thermodynamic susceptibilities, fundamental properties of the QGP medium which are calculable in lattice QCD.  The susceptibilities describe the response of a thermalized system to changes in external conditions and are defined as the partial derivatives of the reduced pressure with respect to the reduced chemical potential, $\hat{\chi}_n^{N=Q,S,B} = \slfrac{\partial^n (P/T^4)}{\partial (\mu_N/T)^n}$, where $Q$, $S$, and $B$ correspond to electric charge, strangeness, and baryon number.  Measurements of net-pion, net-kaon, and net-proton fluctuations have been carried out within ALICE~\cite{AnarQM}, and are related to net-charge, net-strangeness, and net-baryon number fluctuations, respectively.  In this analysis, the fluctuations measurements are extended to the $\Lambda$ and anti-$\Lambda$ baryons.  As the lightest strange baryon, the $\Lambda(\anti{\Lambda})$ gives access to the correlated fluctuations of strangeness and baryon number.  Furthermore, since the resonance contributions to $\Lambda(\anti{\Lambda})$ production are quite different from those to proton and kaon production, a measurement of net-$\Lambda$ fluctuations provides additional information on net-baryon number and net-strangeness fluctuations.

\section{Identity Method for $\Lambda$ baryons}

\begin{SCfigure}
\includegraphics[width=0.35\textwidth]{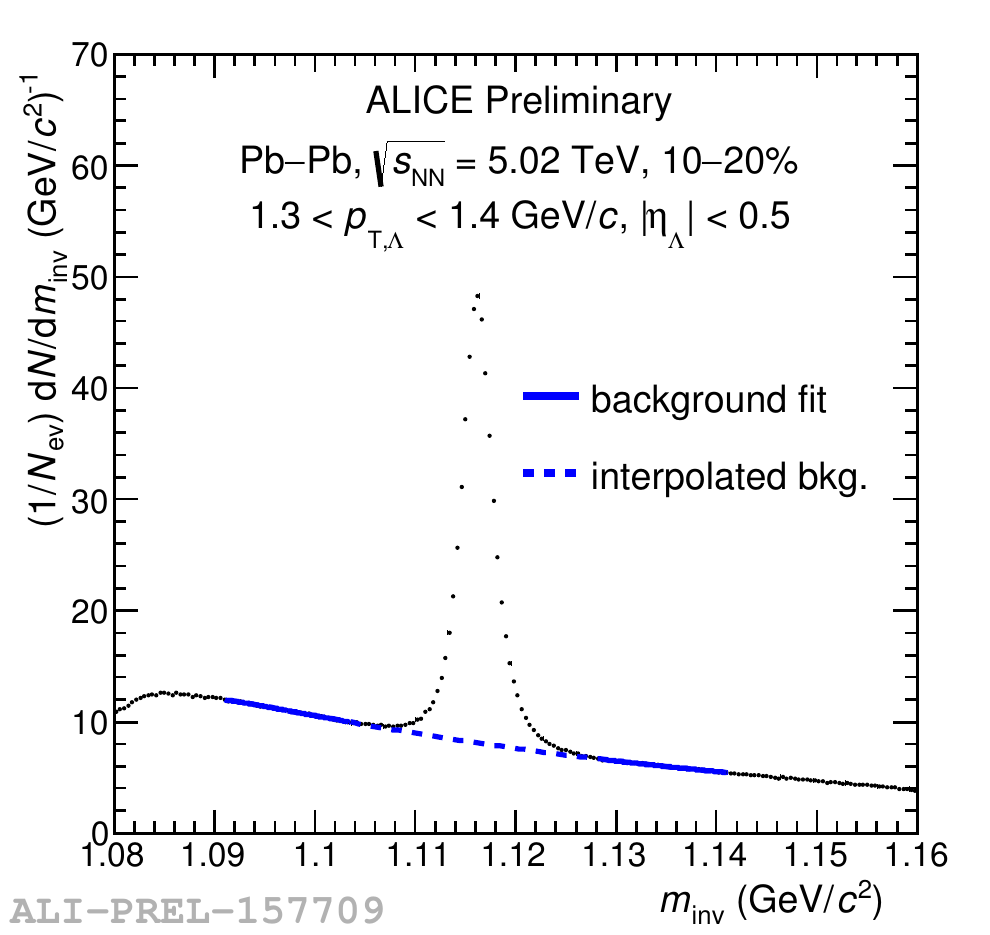}~\includegraphics[width=0.35\textwidth]{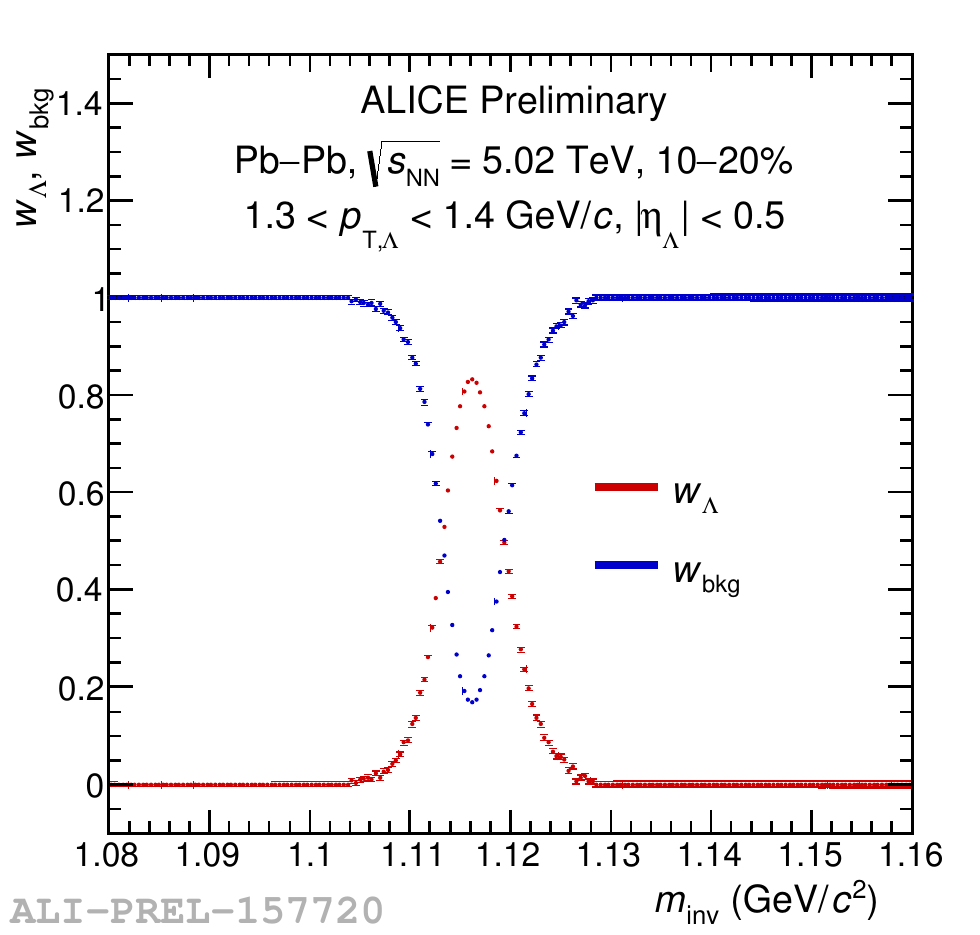}
\caption{(Left) The invariant mass, $m_{inv}$, distribution of $p\pi^{-}$ pairs shows a clear peak around the mass of the $\Lambda$ baryon.  The $m_{inv}$ distribution is fit with a background function in the region outside the peak (solid blue line) and then interpolated to obtain the combinatoric background level underneath the peak (dashed blue line).  (Right) The probabilities that a $p\pi^{-}$ pair is from the decay of a $\Lambda$ (red) or is a combinatoric pair (blue) are shown as a function of $m_{inv}$.\label{fig:exfit}}
\end{SCfigure}

The identification of $\Lambda(\anti{\Lambda})$ baryons proceeds via their decays to protons and pions ($\Lambda\rightarrow p\pi^-$ and $\anti{\Lambda}\rightarrow\anti{p}\pi^+$), as shown in the invariant mass, $m_{inv}$, distribution in Fig.~\ref{fig:exfit}.  However, the presence of the background underneath the $\Lambda(\anti{\Lambda})$ peak in the $m_{inv}$ distribution, due to random combinatorial pairs of protons and pions, makes counting the number of $\Lambda(\anti{\Lambda})$ baryons event-by-event very difficult.  In order to resolve the challenge posed by misidentification in measurements of the higher moments of particle multiplicity distributions, the Identity Method~\cite{ident1,ident2,ident3} was developed.  

Rather than requiring that each particle is identified with absolute certainty, the Identity Method uses information on the \emph{probability} (or `weight,' $w$) that a particle is of given species.  In previous measurements of the moments of the pion, kaon, and proton multiplicity distributions~\cite{AnarQM,MesutQM,MesutPaper}, the weights have been computed as a function of the specific energy loss, $\langle dE/dx\rangle$, of charged particles in the ALICE Time Projection Chamber (TPC).   The Identity Method is then used to account for the momentum ranges in which the $\langle dE/dx\rangle$ distributions for pions, kaons, and protons overlap.  Traditional cut-based analyses~\cite{NirbhayQM} use information from other detectors (for example, the Time-Of-Flight system) or impose strict selection cuts in order to keep the purity of the identified particles high, at the expense of efficiency.  Usage of the Identity Method makes it possible to account for misidentification while keeping the detection efficiency high.  


In the net-$\Lambda$ analysis presented here, $m_{inv}$ is used as the particle identification variable in the Identity Method for the first time.  The probability that a proton-pion pair with a certain invariant mass is the product of a $\Lambda(\anti{\Lambda})$ decay, $w_{\Lambda}$($w_{\anti{\Lambda}}$), or a random combinatoric pair, $w_{bkg}$, can be determined from the inclusive $m_{inv}$ distribution measured with high precision in the full event sample.  These probabilities are evaluated by fitting the combinatoric background in the $m_{inv}$ distribution in the regions away from the $\Lambda$ mass peak, and then interpolating the fit function below the peak to determine the background level, as shown in Fig.~\ref{fig:exfit}.  

Instead of counting the number of particles of a given species (i.e. $N_{\Lambda}$, $N_{\anti{\Lambda}}$) in a single event, in the Identity Method the sum of weights ($W_{\Lambda}$, $W_{\anti{\Lambda}}$) is computed.  For example, 
$W_{\Lambda} = \sum_{i=1}^{N_{p\pi^-}} w_{\Lambda}(m_{inv}^i)$, 
where the summation is performed over all proton-pion pairs in a single event, $m_{inv}^i$ is the invariant mass of the $i^{th}$ pair, and $w_{\Lambda}(m_{inv})$ is the probability that a pair with invariant mass $m_{inv}$ comes from the decay of a $\Lambda$.  The Identity Method provides a mathematical framework for transforming the event-averaged moments of the $W$ distributions ($\langle W^2_{\Lambda}\rangle$, $\langle W^2_{\anti{\Lambda}}\rangle$, $\langle W_{\Lambda}W_{\anti{\Lambda}}\rangle$, etc.) into the moments of the multiplicity distribution: $\langle N^2_{\Lambda}\rangle$, $\langle N^2_{\anti{\Lambda}}\rangle$, $\langle N_{\Lambda}N_{\anti{\Lambda}}\rangle$, etc.  The second moments of the net-$\Lambda$ distribution are then easily calculated as $\langle N_{\Lambda-\anti{\Lambda}}^2\rangle = \langle N_{\Lambda}^2\rangle + \langle N_{\anti{\Lambda}}^2\rangle - 2\langle N_{\Lambda}N_{\anti{\Lambda}}\rangle$.

\section{Corrections and systematic uncertainties}

The probability of reconstructing a given $\Lambda(\anti{\Lambda})$ in the detector is low due to the branching ratio to $p\pi^-(\anti{p}\pi^+)$, which is around 64\%, and because the pion daughter typically has low momentum and may not be reconstructed in the TPC.  This pair reconstruction efficiency, $\varepsilon$, which is evaluated in Monte Carlo (MC) simulations using events from the HIJING~\cite{hijing} event generator processed through a GEANT model of the ALICE detector, varies from 10\% at $\pts{\Lambda} = 1\GeVc$ to 30\% at $\pts{\Lambda} = 4\GeVc$.  Furthermore, 20\% to 35\% of the reconstructed $\Lambda(\anti{\Lambda})$ do not originate from the primary collision vertex, but rather from the decay of $\Xi$ baryons.  The primary $\Lambda$, $\anti{\Lambda}$, and net-$\Lambda$ moments are corrected for the $\pta{}$-dependent reconstruction efficiency and contamination fraction ($\delta$), combined in the factor $\varepsilon/(1-\delta)$, using the procedure described in~\cite{ClaudeIM}.  


The application of the Identity Method along the $m_{inv}$ axis as well as the efficiency and contamination correction procedure were validated in a MC closure test.  The systematic uncertainties on the measurement include the small deviations from the closure test, the uncertainties on $\varepsilon$ and $\delta$ due to the ALICE detector material budget and $\Xi$ spectra, the $m_{inv}$ distribution fitting procedure, variations of the cuts used in the $\Lambda(\anti{\Lambda})$ reconstruction, and collision event pileup rates.  The statistical uncertainties were calculated using the subsample method with $N_{sub}=30$.  

\section{Results}

\setlength{\columnsep}{0.02\textwidth}%
\begin{wrapfigure}{r}{0.47\textwidth}
\centering
\includegraphics[width=0.47\textwidth]{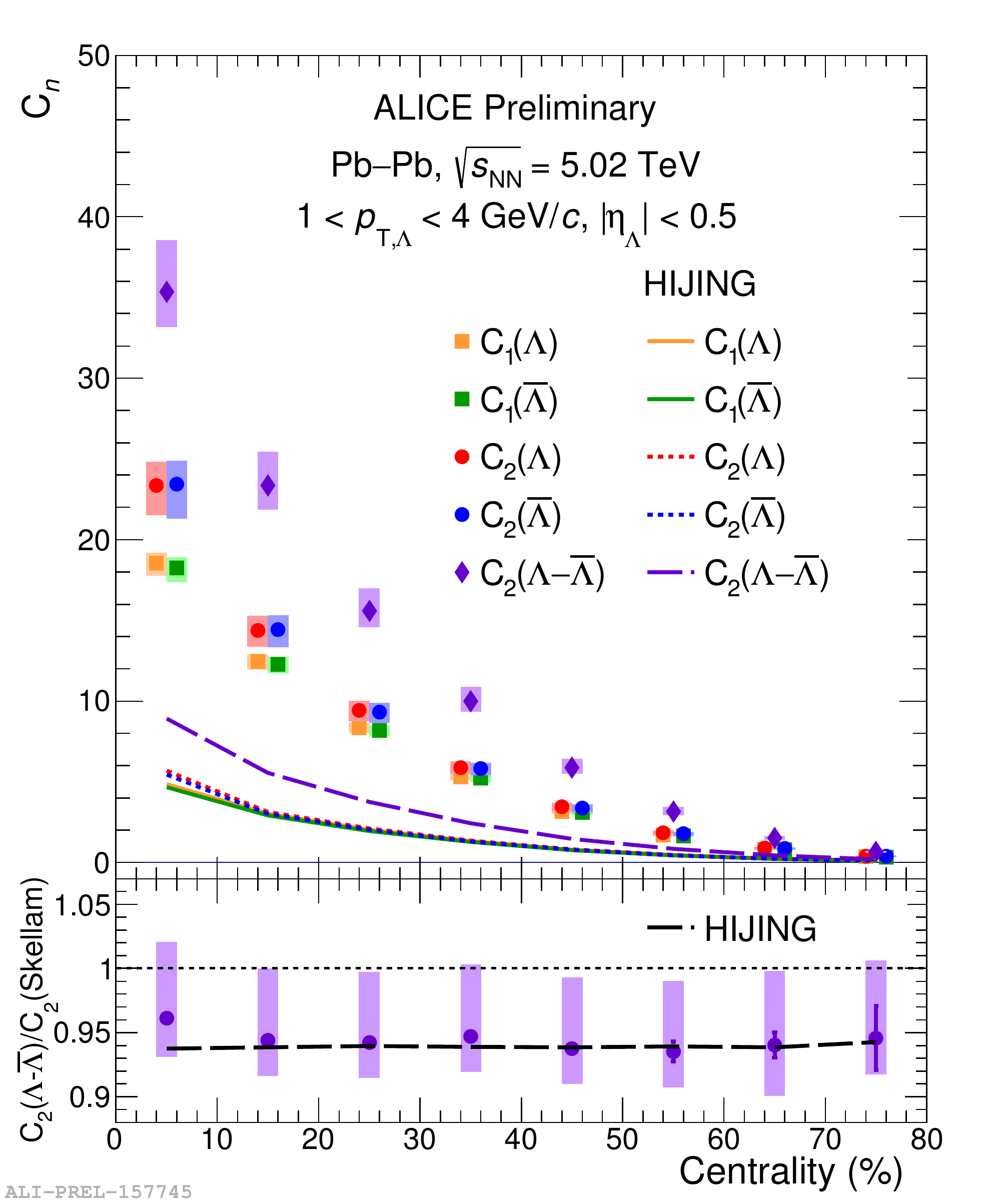}
\caption{(Top) Centrality dependence of the first and second moments of $\Lambda$, $\anti{\Lambda}$, and net-$\Lambda$ fluctuations. (Bottom) Ratio of $C_2(\Lambda-\anti{\Lambda})$ to the Skellam baseline $C_2(\text{Skellam}) = C_1(\Lambda)+C_1(\anti{\Lambda})$.  The results are compared to HIJING predictions.\label{fig:centdep}}
\end{wrapfigure}

The measured first ($C_1$) and second ($C_2$) cumulants of $\Lambda$ and $\anti{\Lambda}$ baryons, as well as the second cumulants of the net-$\Lambda$ distribution ($C_2(\Lambda-\anti{\Lambda})$), are shown as a function of centrality in Fig.~\ref{fig:centdep}.  

If the multiplicity distributions of $N_{\Lambda}$ and $N_{\anti{\Lambda}}$ are Poissonian and uncorrelated, then the resulting distribution of $N_{\Lambda}-N_{\anti{\Lambda}}$ is Skellam.  The higher moments of a Skellam distribution are simply related to the first moments of the original independent Poissonian distributions: $C_n(\text{Skellam}) = C_1(\Lambda)+(-1)^{n}C_1(\anti{\Lambda})$.  The ratio to the Skellam baseline, $C_2(\Lambda-\anti{\Lambda})/\left(C_1(\Lambda)+C_1(\anti{\Lambda})\right)$, is also shown in Fig.~\ref{fig:centdep}, and the results are compared with predictions from the HIJING MC event generator~\cite{hijing}.  The measured ratio of the second moments of the net-$\Lambda$ distribution to the Skellam expectation shows slight indications of a deviation from unity, consistent with observations from the net-proton analysis~\cite{AnarQM}, despite significant systematic uncertainties.  While HIJING describes the trend with centrality well, it significantly underestimates the magnitude of the measured moments (by roughly a factor of four).  HIJING also indicates a deviation from Skellam, although in the case of the MC generator the underlying cause for this deviation is unclear.  

\begin{SCfigure}
\centering
\includegraphics[width=0.38\textwidth]{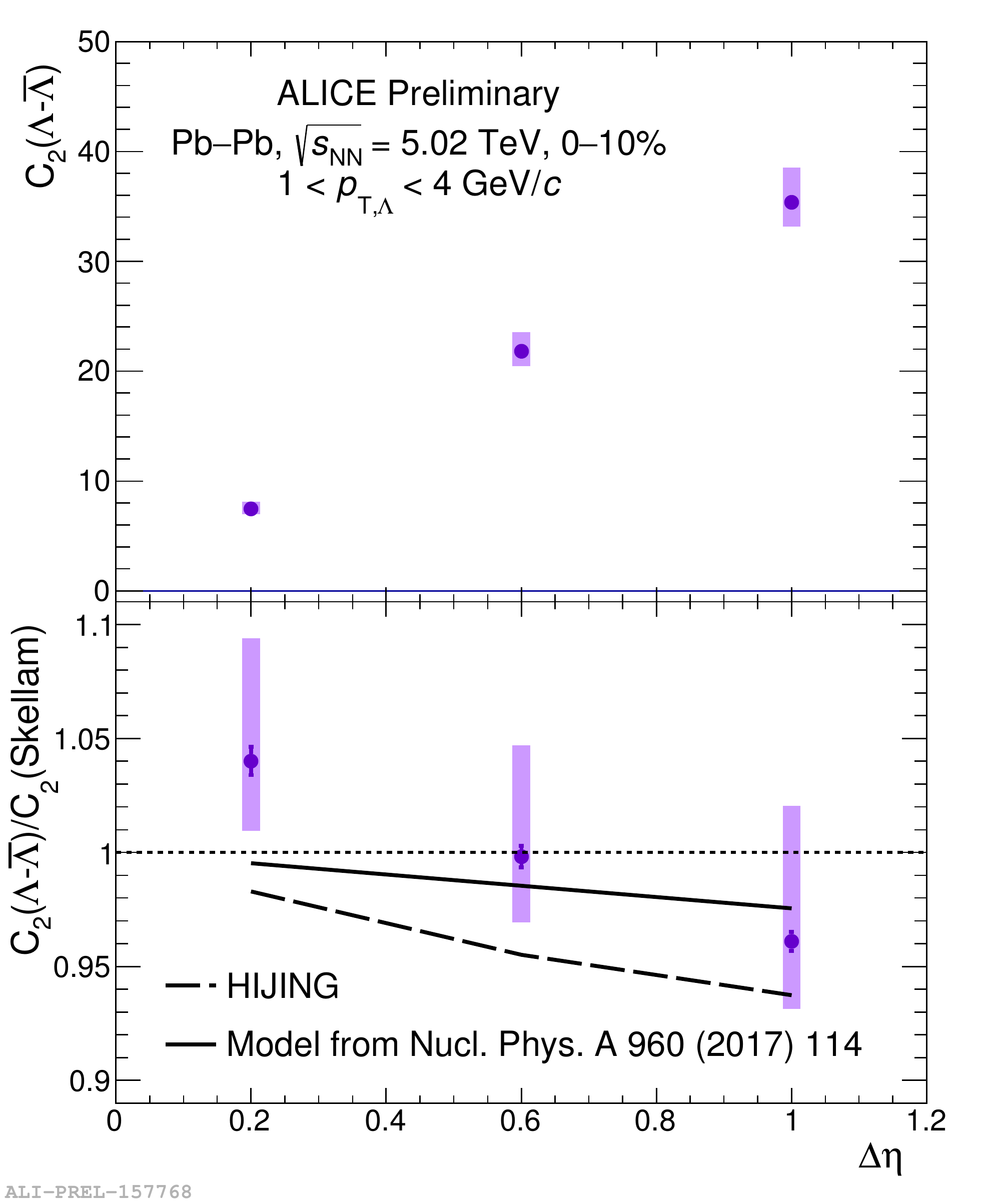}~\includegraphics[width=0.38\textwidth]{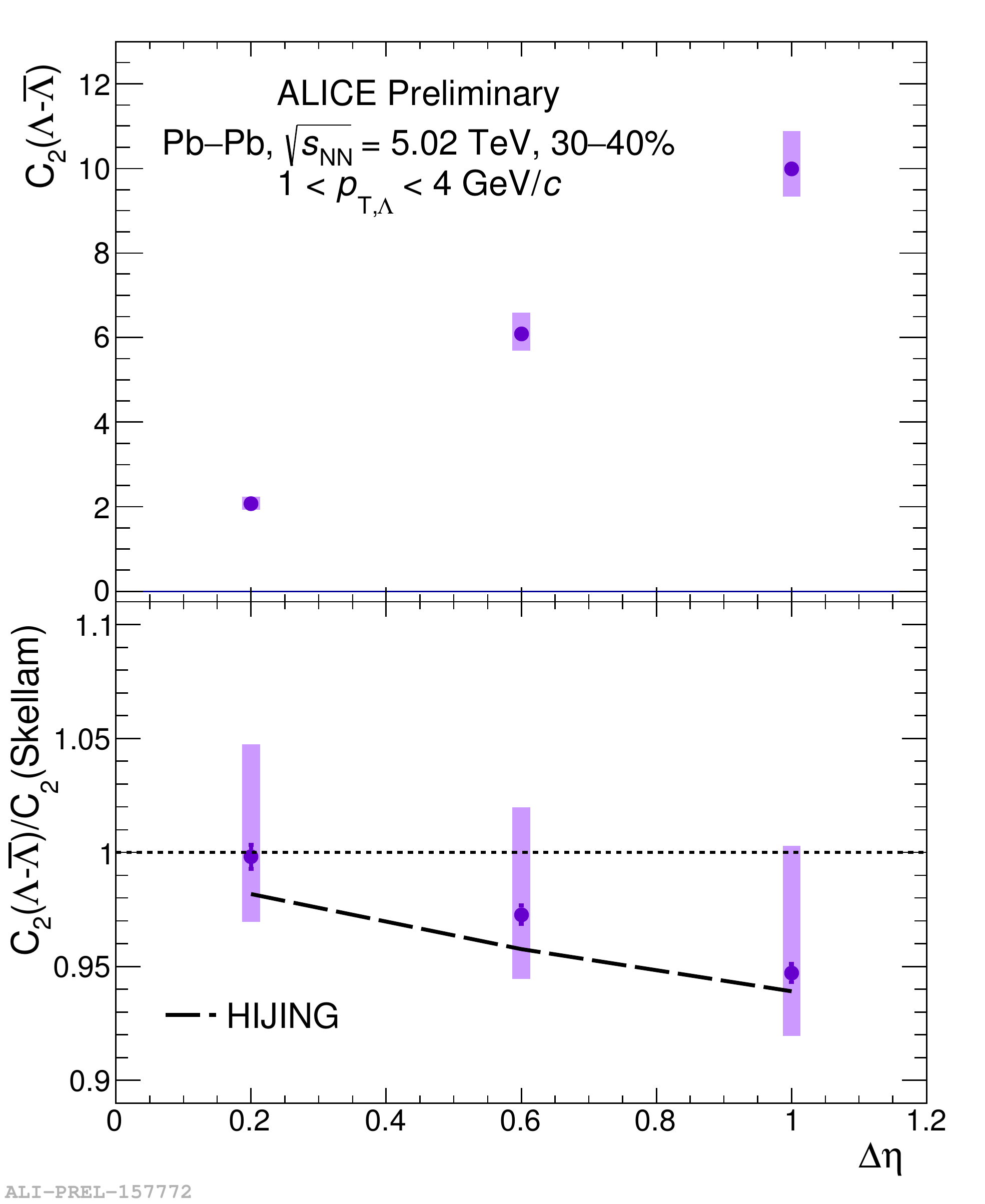}
\caption{(Top) $\Delta\eta$ dependence of the second moments of net-$\Lambda$ fluctuations and (bottom) the ratio to the Skellam baseline in (left) 0-10\% and (right) 30-40\% central events.  The results are compared to expectations from (solid line) a model including the effects of global baryon conservation~\cite{AnarPBMJohanna} and (dashed lines) the HIJING MC generator.  \label{fig:etadep}}
\end{SCfigure}

In the net-proton analysis, the observed small deviation from the Skellam expectation was attributed to global baryon number conservation~\cite{AnarQM}.  The effects of global conservation laws can be tested by exploring the dependence of the moments on the pseudorapidity acceptance, $\Delta\eta$, of the measurement.  One would expect that, if the pseudorapidity acceptance of the measurement is small compared to the pseudorapidity extent of particle production, only Poissonian fluctuations would be present and the effects of global conservation laws would be small.  When the pseudorapidity acceptance of the measurement is large compared to the pseudorapidity distribution of produced particles, then global conservation laws would cause deviations from Poissonian behavior even in the absence of critical fluctuations.  As shown in Fig.~\ref{fig:etadep}, the ratio of $C_2(\Lambda-\anti{\Lambda})$ to the Skellam baseline decreases with increasing $\Delta\eta$ as expected (note that the systematic uncertainties are highly correlated point-to-point).  Such behavior was also observed in the net-proton fluctuations measurement, and is consistent with a model of baryon number conservation effects~\cite{AnarQM,AnarPBMJohanna}, as well as with HIJING expectations.  Future model studies incorporating global strangeness conservation will also provide insight into net-$\Lambda$ fluctuations.  

\section{Summary}

In these proceedings we present the first measurement of the second moments of net-$\Lambda$ fluctuations in Pb--Pb collisions, which have been measured at $\sqrt{s_{\rm NN}} = 5.02~\text{TeV}$ in ALICE.  The centrality dependence of the second moments of the net-$\Lambda$ distribution is shown and compared to results from the HIJING MC generator.  $C_2(\Lambda-\anti{\Lambda})$ decreases from the Skellam baseline with increasing $\Delta\eta$ as is expected from the effects of global conservation laws.  The results of the net-$\Lambda$ analysis are qualitatively consistent with the net-proton fluctuations measurement from ALICE and a model including the effects of global baryon number conservation.  This analysis also represents the first application of the Identity Method to invariant mass distributions.  Therefore this work opens the door for future measurements of higher moments of the multiplicity distributions of strange baryons.

\section*{Acknowledgements}
This work has been supported by BMBF and SFB 1225 ISOQUANT.



\bibliographystyle{elsarticle-num}
\bibliography{aohlson_qm18proceedings}







\end{document}